\PassOptionsToPackage{table}{xcolor}
\documentclass[journal]{IEEEtran}

\newtheorem{rem}{Remark}

\newtheorem{lem}{Lemma}
\newtheorem{prop}{Proposition}

\newtheorem{thm}{Theorem}
\newtheorem{assm}{Assumption}

\newtheorem{claim}{Claim}
\usepackage{cite}
\usepackage{amsmath}
\usepackage{amstext}
\usepackage{amssymb}
\usepackage{tikz}
\usepackage[mathscr]{euscript}

\usepackage{amsfonts}
\usepackage{enumerate}
\usepackage[final]{pdfpages}
\usepackage{csquotes}

\DeclareOldFontCommand{\rm}{\normalfont\rmfamily}{\mathrm}

\makeatletter
\newcommand*{\rom}[1]{\expandafter\@slowromancap\romannumeral #1@}
\makeatother
\usepackage{graphicx}      

\usepackage{xparse}
\usepackage{mathdots}
\usepackage{epstopdf}

\DeclareMathOperator{\diag}{diag}
\definecolor{ForestGreen}{RGB}{34,139,34}
\newcommand{\seb}[1]{%
{\leavevmode\color{black}#1}%
}

\usepackage{array}

\usepackage{mathtools,xparse}

\usepackage{pdfpages}

\usepackage{subcaption}
\usepackage{enumitem}

\def\qed{\hfill $\Box$}

\begin{document}
\title{On the Endemic Behavior of a Competitive  Tri-Virus SIS Networked Model}

\author{Sebin Gracy,
       Mengbin~Ye,
       Brian D.O. Anderson,
        C\'esar A.~Uribe.

\thanks{Sebin Gracy and C\'esar A.~Uribe are with the Department of Electrical and Computer Engineering, Rice University, Houston, TX, USA (\texttt{sebin.gracy@rice.edu}, \texttt{cesar.uribe@rice.edu}).
Mengbin~Ye is with  the School of Electrical Engineering, Computing and Mathematical Sciences, Curtin University, Australia. (
\texttt{mengbin.ye@curtin.edu.au}). Brian D.O. Anderson is with the School of Engineering, Australian National University, Canberra, Australia. (\texttt{brian.anderson@anu.edu.au})}

}

\maketitle
\begin{abstract}
\seb{This paper studies the endemic behavior of a multi-competitive networked susceptible-infected-susceptible (SIS) model. In particular, we focus on the case where there are three competing viruses (i.e., tri-virus system). First, we show that the tri-virus system is not a monotone system. Thereafter, we provide a condition that guarantees local exponential convergence to a boundary equilibrium  (exactly one virus is endemic, the other two are dead), and identify a special case that admits the existence and local exponential attractivity of a line of coexistence equilibria  
(at least two viruses are active). Finally, we identify a particular case (subsumed by the aforementioned special case) such that for all nonzero initial infection levels the dynamics of the tri-virus system converge to a plane of coexistence equilibira.}
\end{abstract}

\maketitle

\section{Introduction}
\seb{Mathematical modeling of spreading 
processes has been an active area of research for several decades. It has spanned multiple scientific disciplines such as physics \cite{van2008virus}, mathematics \cite{hethcote2000mathematics}, economics \cite{bloom2018epidemics}, computer science \cite{prakash2010virus}, etc. The central theme involving all such research directions is to derive a foundational understanding of what causes a disease to spread, and, then, exploit the said understanding to design effective mitigation (or eradication) strategies. Towards this end, various models have been proposed in the literature. This paper deals with the susceptible-infected-susceptible (SIS) model. 
 More specifically, we are interested in networked SIS models,  where each node in the network represents a large population and interconnections between nodes  capture the possibility of the virus spreading between populations \cite{lajmanovich1976deterministic,fall2007epidemiological,khanafer2016stability}.}
\par \seb{The vast majority of the literature on (networked) SIS models concerns the presence of a single virus. However, in practice, one often encounters several strains of a virus that might be simultaneously circulating in a community, as observed during the ongoing COVID-19 crisis.  In a scenario where multiple viruses are present, these viruses could possibly be competing\footnote{\seb{Another possibility is for the multiple viruses to be co-operative; see \cite{gracy2022modeling} for analysis of a special case}.} with each other. That is, assuming there are $m>1$ viruses present, 
an individual (belonging to a population node) can be infected with at most one virus at any given time. More precisely, being infected with one virus precludes the possibility of being simultaneously infected with any of the other $m-1$ viruses.
Other settings where such a competing phenomenon is observed include, but are not limited to, spread of competing opinions on different social networks, competing products in a market, and  spread of conflicting rumors \cite{nedic2019graph}. This motivates the need for multi-competitive networked SIS models.
Analysis of multi-competitive networked SIS  models 
is a much harder problem than in the single virus case, mainly because the dynamics exhibited in the multi-virus setting are far richer than those in the single-virus setting, e.g., there can be multiple attractive equilibria~\cite{castillo1989epidemiological,carlos2,sahneh2014competitive,santos2015bi,liu2019analysis,pare2021multi}. The case where $m=2$ (also referred to as competitive bivirus spread) has been relatively well-explored in recent times; see \cite{sahneh2014competitive,santos2015bi,liu2019analysis,pare2021multi,ye2021convergence,axel2020TAC,ben:brian:opinion:lcss}. However, settings accounting for the presence of more than two competing viruses have not been well studied. The paper \cite{pare2021multi} (also see \cite{pare2020analysis} for the discrete-time version) proposes a multi-competitive networked SIS model, where $m$ (with $m$ being arbitrary but finite) viruses are simultaneously active. However, the analysis of the endemic behavior in \cite{pare2021multi,pare2020analysis} is rather restrictive. The present paper focuses on the special case when $m=3$, i.e., the tri-virus competitive networked SIS model. The set of equilibria of such a system can be broadly classified into three categories: the disease-free equilibrium (all three viruses have been eradicated); the boundary equilibria (two viruses are dead, and one is alive); and coexistence equilibria (at least two viruses infect separate fractions of every population node in the network).}
\seb{Our contributions are as follows:
\begin{enumerate}[label=\roman*)]
    \item We show that the tri-virus system, unlike the bi-virus system, is not a monotone system; see Theorem~\ref{prop:tri-virus-not-monotone}.
    \item We identify a necessary and sufficient condition for local exponential convergence to a boundary equilibrium; see Theorem~\ref{thm:local}. 
    \item \label{contribution:3} We identify a special case (with respect to the class of system paramaters) that admits the existence (and, under certain conditions, guarantees the local exponential attractivity) of a line of coexistence equilibria; see Theorem~\ref{thm:init:condns}. 
    \item  For a special case, subsumed by that in~\ref{contribution:3}, we   provide a sufficient condition which ensures that, regardless of the initial non-zero infection levels, the dynamics of the tri-virus system converge to a plane of coexistence equilibria; see Theorem~\ref{thm:global:plane}.
\end{enumerate}
}
\subsection*{Paper Outline}
\seb{The paper unfolds as follows. We conclude the present section by listing the key notations needed in the rest of the paper. The model, main assumptions, connections between monotone dynamical systems and multi-competitive networked SIS models, and problem statements of interest are detailed in Section~\ref{sec:prob:formulation}. We rigorously show that the tri-virus system is not monotone in Section~\ref{sec:trivirus:monotone}, whereas Section~\ref{sec:persistence:one:virus} deals with the case where at least one virus persists in the network. The analysis pertaining to the existence and  attractivity of a continuum of coexistence equilibria is split across Sections~\ref{sec:line:attractivity} and~\ref{sec:global:plane}. A summary of the paper along with some future directions of possible interest are provided in Section~\ref{sec:conclusion}.}
\subsection*{Notations}
We denote the set of real numbers by $\mathbb{R}$, and the set of nonnegative real numbers by $\mathbb{R}_+$. For any positive integer $n$, we use $[n]$ to denote the set $\{1,2,...,n\}$. The $i^{\rm{th}}$ entry of a vector $x$ is denoted by $x_i$. The element in the $i^{\rm{th}}$ row and $j^{\rm{th}}$ column of a matrix $M$ is denoted by $M_{ij}$. We use $\textbf{0}$ and $\textbf{1}$ to denote the vectors whose entries all equal $0$ and $1$, respectively, and use $I$ to denote the identity matrix, while the sizes of the vectors and matrices are to be understood from the context. For a vector $x$ we denote the square matrix with $x$ along the diagonal by $\diag(x)$. For any two real vectors $a, b \in \mathbb{R}^n$ we write $a \geq b$ if $a_i \geq b_i$ for all $i \in [n]$, $a>b$ if $a \geq b$ and $a \neq b$, and $a \gg b$ if $a_i > b_i$ for all $i \in [n]$. Likewise, for any two real matrices $A, B \in \mathbb{R}^{n \times m}$, we write $A \geq B$ if $A_{ij} \geq B_{ij}$ for all $i \in [n]$, $j \in [m]$, and $A>B$ if $A \geq B$ and $A \neq B$. 
For a square matrix $M$, we use $\sigma(M)$ to denote the spectrum of $M$, $\rho(M)$ to denote the spectral radius of $M$, and $s(M)$ to denote the largest real part among the eigenvalues of $M$, i.e., $s(M) = \max\{\rm{Re}(\lambda) : \lambda \in \sigma(M)\}$.\\ 
A square matrix $A$ is said to be Hurwitz if $s(A)<0$.
A real square matrix $A$ is said to be Metzler if all of its off-diagonal entries are nonnegative. A real square matrix $A$ is said to be a Z-matrix if all of its off-diagonal entries are nonpositive. 
A Z-matrix is an M-matrix if all its eigenvalues have nonnegative real parts. Furthermore, if an M-matrix has an eigenvalue at the origin, then we say that it is singular; if each of its eigenvalues have strictly positive parts, then we say that it is nonsingular. The matrix $A$ is said to be positive semidefinite if $x^\top A x \geq 0$ for all vectors $x$, and we denote this by $A\succeq 0$.

\section{Problem Formulation}\label{sec:prob:formulation}
In this section, we detail a model that captures the spread of multiple competing viruses  across a population network. 
Subsequently, we detail the  pertinent assumptions and definitions that will be required in the sequel. Finally, we formally specify the problems being investigated.

\subsection{Model}
\seb{Consider a network of $n\geq 2$ 
nodes\footnote{As an aside, SIS models where $n=1$ have been studied in, among others, \cite[Section~2]{pare2020modeling}}, where $m$ viruses compete with each other to infect the 
nodes. The notion of competition implies the presence of at least two (but possibly more) viruses. 
Throughout this paper, $m = 3$.
In context, each node 
represents a well-mixed population of individuals with a large and constant size. A well-mixed population means any two individuals in the population can interact with the same positive probability. A key assumption that underpins this model is that of homogeneity within the population node, and (possible) heterogeneity outside the population node. That is, all individuals within a population node have the same infection (resp. healing) rates, but individuals in different population nodes need not necessarily have the same healing (resp. infection) rate\cite{lajmanovich1976deterministic}}.

\seb{Within each population, individuals can be partitioned into four mutually exclusive health compartments: susceptible, infected with virus 1, infected with virus 2, infected with virus~3. More precisely,  no individual can be \emph{simultaneously} infected by more than one virus.
We say a population node is healthy if all individuals belong to the susceptible compartment; otherwise we say it is infected.
An individual, belonging to population $i$ (where $i\in [n]$), in the susceptible compartment, can transition to the ``infected with virus $k$" (for $k \in [m]$) compartment at a rate $\beta_i^k > 0$. An individual in population~$i$ that is infected with virus~$k$ recovers from it 
based on 
\seb{said individual's}
healing rate \seb{with respect to virus~$k$, i.e.,} $\delta_i^k > 0$.}


\par The spread of $m$-competing viruses can be modeled using an $m$-layer graph $G$, where the vertices of the graph represent the population nodes. The $k^\textrm{th}$ layer denotes the contact graph for the spread of virus~$k$, for each $k \in [m]$. More specifically, for the  graph $G$, there exists a directed edge from node $j$ to node $i$ in layer $k$ if, assuming an individual in population $j$ is infected with virus~$k$, then said individual can infect at least one (but possibly more) healthy individual in node~$i$. Let $E^k$ denote the edge set corresponding to the $k^\textrm{th}$ layer of $G$.
We denote by $A^k$ (where $a_{ij}^k \geq 0$) the weighted adjacency matrix corresponding to layer~$k$, with the elements in $A^k$ being in one-to-one correspondence with the existence (or lack thereof) of edges in layer $k$.  That is,  $(i,j)\in E^k$ if, and only if, $a_{ji}^k\neq 0$. Let $x_i^k(t)$ denote the fraction of individuals infected with virus~$k$ in population~$i$ at time instant $t$. The evolution of this fraction can, then, be represented by the following scalar differential equation \cite[Equation~4]{pare2021multi}:
\begin{equation} \label{eq:scalar}
   \dot{x}_i^k(t) = - \delta_i^k x_i^k(t) + \big{(} 1 - \textstyle \sum_{l=1}^m x_i^l(t) \big{)} 
  \textstyle \sum_{j=1}^{n} \beta_{ij}^k x_j^k(t),
  \end{equation}
where $\beta_{ij}^k = \beta_i^ka_{ij}^k$.

Define $x^k(t) = [x_1^k(t), \hdots, x_n^k(t)]^\top$, $D^k =\diag{(\delta_i^k)}$, and $B^k=[\beta_{ij}^k]_{n \times n}$. Therefore, ~\eqref{eq:scalar} can be written as
\begin{equation} \label{eq:vec}
   \dot{x}^k(t) =  \Big{(} - D^k+ \big{(} I - \textstyle \sum_{l=1}^m \diag(x^l(t)) \big{)} B^k  \Big{)} x^k(t), 
  \end{equation}
Defining $x(t):=[x^1(t),  \dots , x^m(t)]^T$,  and  $R^k(x(t)) := \big{(} - D^k + (I - \textstyle \sum_{l=1}^m \diag(x^l(t)))B^k \big{)}$, the dynamics of the system of all $m$ viruses are given by 
\begin{gather} \label{eq:full}
 \dot{x}(t)
 =
 \begin{bmatrix}
 R^1 \big{(} x(t) \big{)} & 0 & \dots & 0 \\
 0 & R^2 \big{(} x(t) \big{)} & \dots & 0 \\
 \vdots & \vdots & \ddots & \vdots \\
 0 & 0 & \dots & R^m \big{(} x(t) \big{)}
 \end{bmatrix}
 x(t).
\end{gather}  
  
Note that by setting $m=1$ in~\eqref{eq:full}, one recovers the classic single-virus SIS model that has been studied extensively in the literature; see \cite{van2008virus,khanafer2016stability,lajmanovich1976deterministic}. Setting $m=2$ yields the classic networked bi-virus SIS model, for which a plethora of results have been provided in \cite{ye2021convergence,liu2019analysis,sahneh2014competitive,santos2015bi,castillo1989epidemiological}.
In this paper, we are interested in the case when $m=3$, i.e., the tri-virus networked competitive spread.


\normalsize
\par Define, for $k \in [3]$, $X^k=\diag(x^k)$. Based on~\eqref{eq:vec}, the dynamics of the tri-virus system can be written as follows:

\begin{align} 
   \dot{x}^1(t) &=  \Big{(} \big{(} I - (X^1+X^2+X^3) \big{)} B^1 - D^1 \Big{)} x^1(t), \label{eq:x1}\\
      \dot{x}^2(t) &=  \Big{(} \big{(} I - (X^1+X^2+X^3) \big{)} B^2 - D^2 \Big{)} x^2(t) \label{eq:x2}\\
         \dot{x}^3(t) &=  \Big{(} \big{(} I - (X^1+X^2+X^3) \big{)} B^3 - D^3 \Big{)} x^3(t). \label{eq:x3}
  \end{align}
 
\subsection{Assumptions and Preliminary Lemmas}  
 We need the following assumptions to ensure that the aforementioned model is well-defined.
 \begin{assm} \label{assum:base}
Suppose that $\delta_i^k>0,  \beta_{ij}^k \geq 0$  for all $i, j \in [n]$ and $k \in [3]$. 
\end{assm}

\begin{assm}\label{assum:irreducible}
The matrix $B^k$, for $ k \in [3]$ is irreducible. \end{assm} 
 
Observe that under Assumption~\ref{assum:base},  for all $k \in [3]$, $B^k$ is a nonnegative matrix and $D^k$ is a positive diagonal matrix. 
Moreover, recall that a square nonnegative matrix $M$ has the irreducibility property if and only if, supposing $M$ is the  (un)weighted adjacency matrix of a graph, the corresponding graph is strongly connected.
Then, noting that non-zero elements in $B^k$ represent directed edges in the set $E^k$, we see that $B^k$ is irreducible whenever the $k^{\rm{th}}$ layer of the multi-layer network $G$ is strongly connected. 

\par  Thanks to 
Assumption~\ref{assum:base}, we can
restrict our analysis to the sets $\mathcal{D} \coloneqq \{x(t): x^k(t) \in [0,1]^n,  \forall k \in [3], \sum_{k=1}^{3}x^k \leq \textbf{1}\}$ and $\mathcal{D}^k \coloneqq \{x^k(t) \in [0,1]^n\}$. 
Since $x_i^k(t)$ is to be interpreted as a fraction of a population, 
these sets represent the 
sensible domain of the 
system. 
That is, 
if $x^k(t)$ takes values outside of $\mathcal{D}^k$, then those values would lack physical meaning. The following lemma shows that $x(t)$ never leaves the set $\mathcal{D}$.
%
%
\begin{lem}{\cite[Lemma~1]{pare2021multi}}\label{lem:pos}
Let Assumption~\ref{assum:base} hold. Then $\mathcal{D}$ is positively invariant with respect to~\eqref{eq:full}.
\end{lem} 


Clearly,  $(\textbf{0}, \textbf{0},\textbf{0})$ is an equilibrium of~\eqref{eq:x1}-\eqref{eq:x3}, and is referred to as the disease-free equilibrium (DFE). 
A sufficient condition for global exponential stability (GES) of the DFE is as follows:
\begin{prop}\cite[Theorem~1]{axel2020TAC}\label{prop:exp:convergence}
Consider system~\eqref{eq:x1}-\eqref{eq:x3} under Assumption~\ref{assum:base}. If $s(-D^k+B^k) < 0$, for each $k \in [3]$, then the DFE  is exponentially stable, with domain of attraction containing~$\mathcal{D}$.
\end{prop}

Clearly, the conditions in Proposition~\ref{prop:exp:convergence} also imply asymptotic convergence to the DFE. However, even when the strict inequalities in the said proposition are relaxed,  one can still achieve asymptotic convergence to the DFE.
This is formalized in the next proposition, which is  a generalization of an analogous result for the bivirus setting (see \cite[Theorem~1]{liu2019analysis}).
\begin{prop}\cite[Lemma~2]{pare2021multi}
Consider system~\eqref{eq:x1}-\eqref{eq:x3} under Assumption~\ref{assum:base}. If $s(-D^k+B^k) \leq 0$, for each $k \in [3]$, 
then the DFE is the unique equilibrium of system~\eqref{eq:x1}-\eqref{eq:x3}. Moreover, it is asymptotically stable with the domain of attraction $\mathcal D$.\label{prop:phil}
\end{prop}



It turns out that for every eigenvalue condition in Proposition~\ref{prop:phil} that is violated, one obtains an equilibrium of the form $(\textbf{0}, \dots, \Tilde{x}^k, \dots, \textbf{0})$ in $\mathcal{D}$, where $\Tilde{x}^k$ is the single-virus endemic equilibrium corresponding to virus~$k$. This is formalized in the following proposition.
\begin{prop} 
\cite[Theorem~2.1]{fall2007epidemiological} \label{prop:necessity}
Consider system~\eqref{eq:x1}-\eqref{eq:x3} under Assumptions~\ref{assum:base} and~\ref{assum:irreducible}. For each $k \in [3]$, 
there exists a unique single-virus endemic equilibrium $(\textbf{0}, \dots, \Tilde{x}^k, \dots, \textbf{0})$ in $\mathcal{D}$, with $\textbf{0} \ll \Tilde{x}^k \ll \textbf{1}$ if, and only if, $s(B^k - D^k) > 0$.

\end{prop}
Analytic methods for computing the single-virus endemic equilibria have been provided in 
\cite[Theorem~5]{van2008virus}.

The equilibria of the form $(\textbf{0}, \dots, \Tilde{x}^k, \dots, \textbf{0})$ are referred to as the \emph{boundary equilibria}. The equilibria of the form $(\bar{x}^1, \bar{x}^2, \bar{x}^3)$, 
where at least $\bar{x}^i$ and $\bar{x}^j$ ($i, j \in [3], i\neq j$) 
are nonnegative vectors with at least one positive entry in each of $\bar{x}^i$ and $\bar{x}^j$ are referred to as \emph{coexistence equilibria}. It turns  out that  such vectors are in fact strictly positive; see\seb{\cite[Lemma~6]{axel2020TAC}}.

Let  $J(x^1,x^2,x^3)$ denote the Jacobian matrix of system~\eqref{eq:x1}-~\eqref{eq:x3} for an arbitrary  point in the state space. Therefore, $J(x^1,x^2,x^3)$ is as given in~\eqref{jacob} below.

\begin{figure*}[h!]
	{\noindent}
\begin{align}\label{jacob}
&J(x^1,x^2,x^3) = \\
&\scriptsize
\begin{bmatrix}
-D^1+(I-X^1-X^2-X^3)B^1-\diag(B^1x^1) & -\diag(B^1x^1)   & -\diag(B^1x^1)  \\
-\diag(B^2x^2) & -D^2+(I-X^1-X^2-X^3)B^2-\diag(B^2x^2)  & -\diag(B^2x^2)\\
 -\diag(B^3x^3)& -\diag(B^3x^3)& -D^3+(I-X^1-X^2-X^3)B^3-\diag(B^3x^3)  \end{bmatrix}\normalsize,\nonumber
\end{align}
\end{figure*}

\subsection{Monotone dynamical systems and competitive bivirus networked SIS models} 


Observe that for the case when $m=2$, system~\eqref{eq:full} is, under Assumption~\ref{assum:irreducible}, monotone \cite[Lemma~3.3]{ye2021convergence} (and, assuming homogeneous recovery rates, also in \cite[Theorem~18]{santos2015bi}). 
 That is, setting $m=2$ for system~\eqref{eq:full}, suppose that $(x_A^1(0), x_A^2(0))$ and $(x_B^1(0), x_B^2(0))$ are two initial conditions in $\textrm{int}(D)$ satisfying i) $x_A^1(0)>x_B^1(0)$ and ii) $x_A^2(0)<x_B^2(0)$. Since the bivirus system is monotone, it follows that, for all $t$, i) $x_A^1(t)\gg x_B^1(t)$ and ii) $x_A^2(t)\ll x_B^2(t)$. 
Further, it has been shown that, for almost all choices of $D^i$, $B^i$, $i=1,2$, system~\eqref{eq:full} has a finite number of equilibria \cite[Theorem~3.6]{ye2021convergence}. Therefore, from \cite[Theorems~2.5 and~2.6]{smith1988systems} (or \cite[Theorem~9.4]{hirsch1988stability}), we know that for almost all initial conditions in $\mathcal D$, system~\eqref{eq:full} with $m=2$ converges to a stable equilibrium point. The set of initial conditions for which said convergence does not occur (in which case it is  either a)  already at an unstable equilibrium, assuming such an equilibrium exists, or b) in the stable manifold of an unstable equilibria, 
or c) on a nonattractive limit cycle, or d) in the stable manifold of a nonattractive limit cycle) has measure zero. It is not known if an analogous statement is true for the case when $m=3$. Consequently, understanding the limiting behavior of tri-virus systems remains open if any of the eigenvalue conditions in Proposition~\ref{prop:phil} are violated.

Further,  for the case when $m=2$, a sufficient condition for local exponential convergence to a boundary equilibrium has been identified in \cite[Theorem~3.10]{ye2021convergence}, whereas for the $m=3$ case no such condition has been identified. Likewise, certain special (nongeneric) scenarios  have been identified which lead to the existence of a continuum of coexistence equilibria; see \cite[Theorems~6 and~7]{liu2019analysis}. Improving upon these results, a broader scenario (but again nongeneric), that accounts for a larger class of parameters, has been identified which leads to not only the existence, but also local exponential attractivity,
of a  continuum of coexistence equilibria; see \cite[Proposition~3.9]{ye2021convergence}. Analogous results for the $m=3$ case are as yet unavailable.

\subsection{Problem Statements} 
Based on the above discussions, our objective in the present paper is to answer the following questions:
\begin{enumerate}[label=\roman*)]
  \item Is the tri-virus system monotone?
    \item Can we identify a sufficient condition for local exponential convergence to a boundary equilibrium? 
    \item Can we identify sufficient condition(s) for the existence and local attractivity of a continuum of coexistence equilibria?
    \item Can we identify a special case(s) where, irrespective of the non-zero initial infection levels, the tri-virus dynamics converge to a continuum of coexistence equilibria?
\end{enumerate}

\section{Is the tri-virus system monotone?} \label{sec:trivirus:monotone}
In this section, we seek to conclusively answer whether (or not) system~\eqref{eq:full} with $m=3$ is monotone.

In order to answer this question, we construct a graph associated with the Jacobian of system~\eqref{eq:x1}-\eqref{eq:x3}, say $\bar{G}$. 
The construction follows the outline provided in \cite{sontag2007monotone}. More specifically,
the graph $\bar{G}$ has $3n$ nodes. The edges of  $\bar{G}$ are based on the entries in the Jacobian matrix $J(x^1, x^2, x^3)$. Specifically, if $[J(x^1, x^2, x^3)]_{ij} <0$ for  $i \neq j$, then we draw an edge labelled with "-" sign;  if  $[J(x^1, x^2, x^3)]_{ij} > 0$ for  $i \neq j$, then we draw an edge labelled with "+" sign. Thus,  $\bar{G}$ is a signed graph. Note that $\bar{G}$ has no self-loops. As an aside, also observe that since $x^k(t)\geq 0$ for $k \in [3]$ and $t\in \mathbb{R}_+$, it is immediate that the sign of the elements in $J(x^1, x^2, x^3)$ do not change with the argument, so that $\bar G$ 
is the same for all points in the interior of $\mathcal D$.  

\par We also need the following concept from graph theory. A signed graph is said to be consistent if every undirected cycle in the graph has a net positive sign, i.e., it has an even number of "-" signs \cite{sontag2007monotone}. We have the following result.
\begin{thm} \label{prop:tri-virus-not-monotone}
System~\eqref{eq:x1}-\eqref{eq:x3} is not monotone.
\end{thm}

\textit{Proof:} Note that the Jacobian $J(x^1, x^2, x^3)$ is a block matrix, with all blocks along the off-diagonal being negative diagonal matrices. Pick any node $i$, where $i \in \{1,2, \hdots, n\}$. Observe that, since  all blocks along the off-diagonal of $J(x^1, x^2, x^3)$ are negative diagonal matrices, it is clear that there exists an edge from node $i$ to node $i+n$, an edge from node $i+n$ to node $i+2n$, and an edge from node $i+2n$ to node $i$. Furthermore, each of these edges have a "-" sign. Hence, a loop starting from node $i$, traversing through nodes $i+n$, $i+2n$ and back to node $i$ is a 3-length cycle that has an odd number of negative signs. Therefore, from \cite[page 62]{sontag2007monotone}, the signed graph $\bar{G}$ is not consistent. 
Consequently, from \cite[page 63]{sontag2007monotone}, it follows that the system~\eqref{eq:x1}-\eqref{eq:x3} is not monotone.~\qed


\par Theorem~\ref{prop:tri-virus-not-monotone} sheds light on a very interesting phenomenon, namely that the tri-virus system is not monotone. This is in sharp contrast to the bi-virus setting, which is known to be monotone \cite{ye2021convergence}. The fact that a bivirus system is monotone coupled with the fact that for almost all choices of $D^k$, $B^k$, $k=1,2$, the bivirus system has a finite number of equilibria allows one to draw general conclusions on the limiting behavior of bivirus dynamical systems. 
By extending the algebraic geometry arguments in the proof of \cite[Theorem~3.6]{ye2021convergence}, it is relatively straightforward to show that even for the tri-virus system, for almost all choices of $D^k$, $B^k$, $k=1,2,3$, there exists a finite number of equilibria. The details are omitted here in the interest of space. Nonetheless, due to the findings of Theorem~\ref{prop:tri-virus-not-monotone},
one cannot draw upon the rich literature on monotone dynamical systems (see\cite{smith1988systems}) to study the limiting behavior of system~\eqref{eq:x1}-\eqref{eq:x3}. In general, for non-monotone systems, no dynamical behavior, including chaos, can be definitively ruled out 
\cite{sontag2007monotone}.

Another possible consequence of the lack of monotonicity is as follows: It is  known that setting $D^k=I$ for $k \in [2]$ has no bearing on either the location of equilibria of system~\eqref{eq:full}
with $m=2$ nor on their (local) stability properties \cite[Lemma~3.7]{ye2021convergence}. That is, for a) bivirus systems, where $D^k=I$ and $B^k = (D^k)^{-1}B^k$ for $k \in [2]$, and b) bivirus systems, where $D^k$s are arbitrary positive diagonal matrices, the location of equilibria are the same for both bivirus systems a) and b). Plus, local stability of an equilibrium in  bivirus system a) implies, and is implied by,  that in bivirus system b).
For system~\eqref{eq:full} with $m=3$, by extending the arguments from \cite[Lemma~3.7]{ye2021convergence},  it is straightforward to show that the \emph{location} of the equilibria is the same when, for $k\in [3]$, $D^k=I$ and $B^k = (D^k)^{-1}B^k$,  and when $D^k$ (resp. $B^k$) are arbitrary positive diagonal (resp. nonnegative) matrices with the $D^k$s not necessarily being equal to each other. However, since the tri-virus system is not monotone, the arguments for stability of equilibria in the proof of \cite[Lemma~3.7]{ye2021convergence} cannot be adapted. As such, for the tri-virus case, preservation of stability properties remains an open question when the healing rates for all nodes with respect to all viruses are `scaled' in the manner above to become unity.

\section{Persistence of one or more viruses}\label{sec:persistence:one:virus}
If one or more of the eigenvalue conditions in Proposition~\ref{prop:phil} is violated, then at least one of the viruses persists in the population.  
This, in turn, gives rise to a richer possible set of behaviors, as we will see in the rest of this paper. 

In this section, we identify a sufficient condition for local exponential convergence to a boundary equilibrium. While similar results exist for the case when $m=2$ (see \cite[Theorem~3.10]{ye2021convergence}), to the best of our knowledge, no such result exists for the $m=3$ case. The following theorem addresses this gap, and establishes that the local stability (resp. instability) of a boundary equilibrium corresponding to 
\seb{virus~1
is dependent on whether (or not) the state matrices, obtained by linearizing the dynamics of viruses~$2$ and~$3$ around the single-virus endemic equilibrium of virus~$1$, are Hurwitz}.


\begin{thm}\label{thm:local}
Consider system~\eqref{eq:x1}-\eqref{eq:x3} under Assumptions~\ref{assum:base} and~\ref{assum:irreducible}. 
The boundary equilibrium $(\Tilde{x}^1, \textbf{0}, \textbf{0})$ is locally exponentially 
stable if, and only if, 
each of the following conditions are satisfied: 
\begin{enumerate}[label=\roman*)]
    \item $\rho((I-\tilde{X}^1)(D^2)^{-1}B^2)<1$; and
    \item $\rho((I-\tilde{X}^1)(D^3)^{-1}B^3)<1$.
\end{enumerate}
If $\rho((I{-}\tilde{X}^1)(D^2)^{-1}B^2)>1$ or if \mbox{$\rho((I{-}\tilde{X}^1)(D^3)^{{-}1}B^3)>1$}, then $(\Tilde{x}^1, \textbf{0}, \textbf{0})$ is unstable.
\end{thm}
The proof follows the strategy outlined for the $m=2$ case in~\cite[Theorem~3.10]{ye2021convergence}.\\
\textit{Proof:} Consider the equilibrium point $(\Tilde{x}^1, \textbf{0}, \textbf{0})$, and note that the Jacobian evaluated at this point is as follows:
\begin{align}
&J(\Tilde{x}^1,\textbf{0}, \textbf{0}) = \\
&\scriptsize
\begin{bmatrix}
-D^1+(I-\Tilde{X}^1)B^1-\hat{B}^1  & {-}\hat B^{1}   & {-}\hat B^{1} \\
 \textbf{0} & {-}D^2{+}(I{-}\Tilde{X}^1)B^2 & \textbf{0} \\
\textbf{0}  & \textbf{0} & {-}D^3{+}(I{-}\Tilde{X}^1)B^3 \end{bmatrix}\normalsize,\nonumber
\end{align}
where $\hat{B}^i=\diag(B^i\Tilde{x}^i)$, for $i=1,2,3$.\\
Observe that the matrix $J(\Tilde{x}^1,\textbf{0}, \textbf{0})$ is block upper triangular. Hence, it is Hurwitz if, and only if,  the blocks along the diagonal are Hurwitz. We will now show that this condition is fufilled, as a consequence of the assumptions of Theorem~\ref{thm:local}.
\par Since $(\Tilde{x}^1, \textbf{0}, \textbf{0})$ is an equilibrium point of system~\eqref{eq:x1}-\eqref{eq:x3}, by considering the equilibrium version of equation~\eqref{eq:x1}, we have the following:
\begin{align}\label{eq:eqm:version:1}
    (-D^1+(I-\Tilde{X}^1)B^1)\tilde{x}^1=\textbf{0}.
\end{align}
By Assumption~\ref{assum:base}, we have that $D^1$ is positive diagonal and $B^1$ is nonnegative. Furthermore, by assumption we know that $B^1$ is irreducible. Moreover, from \seb{\cite[Lemma~6]{axel2020TAC}}, 
it follows that $(I-\Tilde{X}^1)$ is positive diagonal, implying that $(I-\Tilde{X}^1)B^1$ is nonnegative irreducible. Thus, we can conclude that the matrix $(-D^1+(I-\Tilde{X}^1)B^1)$ is irreducible Metzler. 
From \seb{\cite[Lemma~6]{axel2020TAC}}
we know that $\textbf{0} \ll \tilde{x}^1$. Hence, by applying \seb{\cite[Lemma~2.3]{varga1999matrix}} 
to~\eqref{eq:eqm:version:1}, it must be that $\tilde{x}^1$ is, up to a scaling, the only eigenvector of $(-D^1+(I-\Tilde{X}^1)B^1)$ with all entries being strictly positive. Furthermore, $\tilde{x}^1$ is the eigenvector that is associated with, and only with, $s(-D^1+(I-\Tilde{X}^1)B^1)$. Therefore,   $s(-D^1+(I-\Tilde{X}^1)B^1)=0$.\\
Define $Q:=D^1-(I-\Tilde{X}^1)B^1$, and note that $Q$ is an M-matrix. Since $s(-Q)=0$ and $B^1$ is irreducible, it follows that $Q$ is a singular irreducible M-matrix. Observe that $\hat{B}^1$ is a nonnegative matrix, and because $B^1$ is irreducible and $\Tilde{x}^1 \gg \textbf{0}$, it must be that at least one element in $\hat{B}^1$ is strictly positive. Therefore, from \cite[Lemma~4.22]{qu2009cooperative}, it follows that $Q+\hat{B}^1$ is an irreducible non-singular M-matrix, which from \cite[Section~4.3, page~167]{qu2009cooperative} implies that $-Q-\hat{B}^1$ is Hurwitz. Therefore, we have that $s(-D^1+(I-\Tilde{X}^1)B^1-\hat{B}^1)<0$.
\par By assumption, $\rho((I-\tilde{X}^1)(D^2)^{-1}B^2)<1$ and $\rho((I-\tilde{X}^1)(D^3)^{-1}B^3)<1$. Therefore, by noting that $D^2$ (resp. $D^3$) are positive diagonal matrices and $B^2$ (resp. $B^3$) are nonnegative irreducible matrices,
from 
\seb{\cite[Proposition~1]{liu2019analysis}} it follows that $s(-D^2+(I-\tilde{X}^1)B^2)<0$ (resp. $s(-D^3+(I-\tilde{X}^1)B^3)<0$). Since each diagonal block of $J(\Tilde{x}^1,\textbf{0}, \textbf{0})$ is Hurwitz, it is clear that $J(\Tilde{x}^1,\textbf{0}, \textbf{0})$ is Hurwitz. Local 
exponential 
stability of $(\Tilde{x}^1,\textbf{0}, \textbf{0})$, then, follows from \cite[Theorem 4.15 and Corollary~4.3]{khalil2002nonlinear}. 
\par \seb{The proof of necessity follows by first noting that if either condition in statement~i) or that in statement~ii) is violated, then, since at least one of the blcoks along the diagonal of $J(\Tilde{x}^1,\textbf{0}, \textbf{0})$ is not Hurwitz, the  the matrix $J(\Tilde{x}^1,\textbf{0}, \textbf{0})$ is not Hurwitz. Then, by invoking the necessity part of \cite[Theorem 4.15 and Corollary~4.3]{khalil2002nonlinear}, the result follows.}
\par The claim for instability can be proved by noting that if either of the eigenvalue conditions are violated, then, since  $J(\Tilde{x}^1,\textbf{0}, \textbf{0})$ is block diagonal, the matrix  $J(\Tilde{x}^1,\textbf{0}, \textbf{0})$ is not Hurwitz. The result follows from \cite[Theorem~4.7, item ii)]{khalil2002nonlinear}.~\qed
\par Analogous results for the boundary equilibria $(\textbf{0}, \Tilde{x}^2, \textbf{0})$ and $(\textbf{0}, \textbf{0},\Tilde{x}^3)$ can be similarly obtained.

\section{Existence and attractivity of a continuum of equilibria for nongeneric tri-virus networks}\label{sec:line:attractivity}

\par Proposition~\ref{prop:necessity} and Theorem~\ref{thm:local}, respectively, deal with the existence and local exponential convergence to a boundary equilibrium. 
In this section, we are interested in identifying a scenario which guarantees the existence and local exponential attractivity of a continuum of coexistence  equilibria.

Let $z$ denote the single-virus endemic equilibrium corresponding to virus~1, with $Z=\diag(z)$. Therefore, assuming $D^1=I$, the vector $z$ fulfils the following:
\begin{equation}\label{eq:z}
    -I+((I-Z)B^1)z=\textbf{0}.
\end{equation}
Furthermore, since $z$ is an endemic equilibrium, from \seb{\cite[Lemma~6]{axel2020TAC}} 
it follows that $\textbf{0} \ll z \ll \textbf{1}$. Let $C$ be any nonnegative irreducible matrix for which $z$ is also an eigenvector corresponding to eigenvalue one. That is, $Cz=z$. Therefore, from\seb{
\cite[Theorem~2.7]{varga1999matrix}}, 
it follows that $\rho(C)=1$, and that the vector $z$, up to a scaling, is the unique eigenvector of $C$ with all entries being strictly positive. Define
\begin{equation}\label{eq:B2}
    B^2:=(I-Z)^{-1}C.
\end{equation}
We have the following result.
\begin{thm}\label{thm:init:condns}
Consider system~\eqref{eq:x1}-\eqref{eq:x3} under Assumption~\ref{assum:base}. Suppose that $D^k=I$ for $k \in [3]$.
    Suppose that 
    $B^1$ and $B^3$ are arbitrary nonnegative irreducible matrices; and
    vector $z$ and matrix $B^2$ are as defined in~\eqref{eq:z} and~\eqref{eq:B2}, respectively. Then, a set of equilibrium points of the trivirus equations is given by $(\beta_1z, (1-\beta_1)z, \textbf{0})$ for all $\beta_1 \in [0,1]$. Furthermore, \begin{enumerate} [label=\roman*)]
    \item if $s(-I+(I-Z)B^3)<0$, then 
the equilibrium set $(\beta_1z, (1-\beta_1)z, \textbf{0})$, with $\beta_1 \in [0,1]$, is locally exponentially attractive.
    \item if $s(-I+(I-Z)B^3)>0$, then 
the equilibrium set $(\beta_1z, (1-\beta_1)z, \textbf{0})$, with $\beta_1 \in [0,1]$, is unstable.
\end{enumerate}
\end{thm}

\textit{Proof:} We first show that, for all $\beta_1 \in [0,1]$, the  point $(\beta_1z, (1-\beta_1)z, \textbf{0})$ fulfils the equilibrium version of equations~\eqref{eq:x1}-\eqref{eq:x3}. To this end, observe that the right hand side of~\eqref{eq:x1}-\eqref{eq:x3} evaluated at  $(\beta_1z, (1-\beta_1)z, \textbf{0})$ yields:
\begin{align}
    &(-I+(I-\beta_1Z-(1-\beta_1)Z)B^1)\beta_1z \nonumber \\
    &=(-I+(I-Z)B^1)\beta_1z =0, \label{beta1z=0}
\end{align}
where~\eqref{beta1z=0} follows by noting that $\beta_1$ is a scalar, and  $z$ is the single-virus endemic equilibrium corresponding to virus~1. Similarly, 
\begin{align}
    &(-I+(I-\beta_1Z-(1-\beta_1)Z)B^2)(1-\beta_1)z \nonumber \\
    &=(-I+(I-Z)B^2)(1-\beta_1)z \nonumber \\
     &=(-I+(I-Z)(I-Z)^{-1}C)(1-\beta_1)z \nonumber \\
     &=(-I+C)(1-\beta_1)z=0, \label{beta2z=0}
\end{align}
where~\eqref{beta2z=0} follows by noting that $Cz=z$. Thus, from~\eqref{beta1z=0} and~\eqref{beta2z=0}, it is clear that, for every $\beta_1 \in [0,1]$,  $(\beta_1z, (1-\beta_1)z, \textbf{0})$ is an equilibrium point of system~\eqref{eq:x1}-\eqref{eq:x3}; i.e. there is a set of equilibrium points $(\beta_1z, (1-\beta_1)z, \textbf{0})$ with $\beta_1 \in [0,1]$. 
\par Next, observe that, for any $\beta_1 \in [0,1]$, the Jacobian evaluated at $(\beta_1z, (1-\beta_1)z, \textbf{0})$ is as given in~\eqref{jacobian:thm2}.
\begin{figure*}[h]
	{\noindent}
\begin{align} \label{jacobian:thm2}
&J(\beta_1z, (1{-}\beta_1)z, \textbf{0}) =
\scriptsize\begin{bmatrix}
-I+(I-Z)B^1-\diag(B^1\beta_1z) & -\diag(B^1\beta_1z)   & -\diag(B^1\beta_1z)  \\
-\diag(B^2(1-\beta_1)z) & -I+(I-Z)B^2-\diag(B^2(1-\beta_1)z)  & -\diag(B^2(1-\beta_1)z)\\
 \textbf{0}&  \textbf{0}& -I+(I-Z)B^3 \end{bmatrix}\normalsize
\end{align}
\end{figure*}
Hence, we can rewrite $J(\beta_1z, (1-\beta_1)z, \textbf{0})$ as 
\begin{align}\label{J:forhatx1:partitioned:1}
    J(\beta_1z, (1-\beta_1)z, \textbf{0})
    =
    & \scriptsize
    \begin{bmatrix} 
    \bar{J}(\beta_1z, (1{-}\beta_1)z, \textbf{0}) 
      && \hat{J} \\
      \mathbf{0} && {-}I{+}(I{-}Z)B^3
    \end{bmatrix}, 
\end{align}
\normalsize
where
\begin{align}\label{jacob:hatx1:22submatrix:1} 
&\bar{J}(\beta_1z, (1-\beta_1)z, \textbf{0})  \nonumber\\ 
& = \scriptsize
\begin{bmatrix} 
{-}I{+}(I{-}Z)B^1{-}\diag(B^1\beta_1z) & -\diag(B^1\beta_1z)   \\
-\diag(B^2(1-\beta_1)z)& {-}I{+}(I{-}Z)B^2{-}\diag(B^2(1{-}\beta_1)z)
 \end{bmatrix}, \end{align}
 \normalsize
while $\hat{J} =  \scriptsize \begin{bmatrix}-\diag(B^1\beta_1z) \\ -\diag(B^2(1-\beta_1)z)\end{bmatrix}$.
\par Note that the matrix $J(\beta_1z, (1-\beta_1)z, \textbf{0})$ is block upper triangular. Hence, it is clear that $s(J(\beta_1z, (1-\beta_1)z, \textbf{0})) =\max\{s(\bar{J}(\beta_1z, (1-\beta_1)z, \textbf{0}), s(-I+(I-Z)B^3)\}$. \\

\textit{Proof of statement~i):} Consider the matrix \break $\bar{J}(\beta_1z, (1-\beta_1)z, \textbf{0})$.  Define $P: = \begin{bmatrix} I_n && \textbf{0} \\ \textbf{0} && -I_n \end{bmatrix}$. Therefore, 
\begin{align}
    &P\bar{J}((\beta_1z, (1-\beta_1)z, \textbf{0})P \nonumber \\
    & = \scriptsize
\begin{bmatrix} 
{-}I{+}(I{-}Z)B^1{-}\diag(B^1\beta_1z) & \diag(B^1\beta_1z)   \\
\diag(B^2(1-\beta_1)z) & {-}I{+}(I{-}Z)B^2{-}\diag(B^2(1{-}\beta_1)z)
\end{bmatrix}.
\end{align}

Note that $P\bar{J}((\beta_1z, (1-\beta_1)z, \textbf{0})P$ is irreducible Metzler.  Hence, by considering the element-wise positive vector $\begin{bmatrix}z^\top &&z^\top\end{bmatrix}^\top$, and by invoking the equilibrium version of the  equation of the single-virus system corresponding to virus~1, it follows that $P\bar{J}((\beta_1z, (1-\beta_1)z, \textbf{0})Pz=\textbf{0}$. Therefore, from \seb{\cite[Lemma~2.3]{varga1999matrix}},
it follows that $s(P\bar{J}(\beta_1z, (1-\beta_1)z, \textbf{0})P)=0$, which implies $s(\bar{J}(\beta_1z, (1-\beta_1)z, \textbf{0})=0$. Consequently, since, by assumption, $s(-I+(I-Z)B^3)<0$, we obtain
$s(J(\beta_1z, (1-\beta_1)z, \textbf{0}))=0$.
Furthermore, since $s(P\bar{J}(\beta_1z, (1-\beta_1)z, \textbf{0})P)=0$, then since $P\bar{J}(\beta_1z, (1-\beta_1)z, \textbf{0})P$ is irreducible Metzler, 
\seb{by \cite[Theorem~2.7]{varga1999matrix}}
we have that the matrix $\bar{J}(\beta_1z, (1-\beta_1)z, \textbf{0})$  has exactly one eigenvalue at the origin, and all other eigenvalues have negative real parts. Therefore, since $s(-I+(I-Z)B^3)<0$, the matrix $J(\beta_1z, (1-\beta_1)z, \textbf{0})$ has exactly one eigenvalue at the origin, and all other eigenvalues have negative real parts. Hence,  the trivirus equations associated with the line of equilibria define a one-dimensional center manifold along which the Jacobian is singular. Therefore, from \cite[Theorem~8.2]{khalil2002nonlinear} it follows that the set of equilibrium points $(\beta_1z, (1-\beta_1)z, \textbf{0})$, with $\beta_1 \in [0,1]$, is locally exponentially attractive.

\textit{Proof of statement~ii):} By assumption, $s(-I+(I-Z)B^3)>0$. Hence, $s(J(\beta_1z, (1-\beta_1)z, \textbf{0}))>0$. Therefore, from \cite[Theorem~4.7, statement ii)]{khalil2002nonlinear} it follows that set of equilibrium points $(\beta_1z, (1-\beta_1)z, \textbf{0})$, with $\beta_1 \in [0,1]$, is unstable.~$\blacksquare$

Observe that the key idea behind Theorem~\ref{thm:init:condns} is to fix the matrix $B^1$, and then choose $B^2$ (as given in~\eqref{eq:B2}) so as to obtain a locally exponentially attractive (resp. unstable) equilibrium set, namely $(\beta_1z, (1-\beta_1)z, \textbf{0})$, with $\beta_1 \in [0,1]$.
\seb{Clearly, the same idea can be applied by the other two possible pairs, namely $(B^1, B^3)$ and $(B^2, B^3)$, to obtain corresponding locally exponentially attractive (resp. unstable) equilibrium set.}

The findings of Theorem~\ref{thm:init:condns} are not in conflict with the claim for finiteness of equilibria presented in Section~\ref{sec:trivirus:monotone}. 
The elements in matrices $D^i$, $B^i$, $i=1,2,3$, are either a priori fixed to a specific value or they are not. In case of the latter, we refer to those as free parameters, in the sense that these are allowed to take any value in $\mathbb{R}_{+}$. The dimension of the space of free parameters equals the number of free parameters in the tri-virus system.
Each choice of free parameters yields a realization of system~\eqref{eq:x1}-\eqref{eq:x3}. The set of choices of free parameters that fall within the special case identified by Theorem~\ref{thm:init:condns} has measure zero.

\begin{rem}\label{rem:zero:init:condns}
Note that Theorem~\ref{thm:init:condns} admits a non-zero 
initial infection level in the network for virus~3. 
If one were to assume that 
$x_i^3(0) =0$ for all $i \in [n]$, then the trivirus system of equations (i.e., \eqref{eq:x1}-\eqref{eq:x3})  collapses into the bivirus equation set (i.e., equation~\eqref{eq:full} where $m=2$). Under such a setting, Theorem~\ref{thm:init:condns} coincides with \cite[Proposition~3.9]{ye2021convergence}, which, in turn, subsumes \cite[Theorems~6 and~7]{liu2019analysis}; both with respect to i) admitting a larger class of parameters than \cite[Theorems~6 and~7]{liu2019analysis} (and, assuming the setup in \cite{pare2021multi} is restricted to the bi-virus case,  \cite[Corollaries~2 and~3]{pare2021multi}), and ii) providing guarantees for local exponential attractivity.~\qed
\end{rem}

\section{Global convergence to a plane of coexistence equilibria}\label{sec:global:plane}
Section~\ref{sec:line:attractivity} dealt with the existence and attractivity (resp. instability) of a line of coexistence equilibria. Moving beyond this, it is of natural interest to 
identify scenario(s) where a plane of coexistence equilibria could exist, and furthermore, seek condition(s) that guarantee local (resp. global) stability of such a plane of coexistence equilbria. The present section deals with this issue.

We consider a case where three identical copies of a virus are spreading over the same graph as formalized next.

\begin{assm}\label{assm:hetero:samegraph}
We suppose that
\begin{enumerate}[label=(\roman*)]
    \item All three viruses are spreading over the same graph.
    \item For all $i \in [n]$ $\delta_i^1 =\delta_i^2 =\delta_i^3>0$.
    \item For all $i=j \in [n]$ and $(i,j) \in \mathcal E$, $\beta_{ij}^1=\beta_{ij}^2=\beta_{ij}^3$.
\end{enumerate}
\end{assm}

Note that for the  special case identified in Assumption~\ref{assm:hetero:samegraph}, assuming that the setting in \cite{pare2021multi} is restricted to the tri-virus case, the existence of a plane of coexistence equilibrium has been secured by \cite[Corollary~3]{pare2021multi}. However, \cite[Corollary~3]{pare2021multi} does not provide guarantees for even local (let alone global)  convergence to the said plane. \seb{To address this shortcoming, first consider the system}

\begin{equation}\label{eq:positive_linear_x}
\dot x(t) = (-D+(I-\diag(\tilde x))B)x(t),
\end{equation}
where $\tilde x$ is the unique endemic equilibrium of the single virus SIS system associated with $(D, B)$, and with $B$ irreducible. The matrix $Q: = D-(I-\diag(\tilde x))B$ is a singular irreducible $M$-matrix, with a simple eigenvalue at $0$ and all other eigenvalues have positive real part. Associated with this simple zero eigenvalue is the right eigenvector $\tilde x \gg {\bf 0}$ and a left eigenvector $\tilde u^\top \gg {\bf 0}^\top$. We assume that $\tilde u^\top$ is normalised to satisfy $\tilde u^\top \tilde x = 1$. Moreover, there exists a positive diagonal matrix $P$ such that $\bar Q := PQ+Q^\top P \succeq 0$. In fact, if we choose $P = \diag(\tilde u_1/\tilde x_1, \hdots, \tilde u_n/\tilde x_n)$, then it can be verified that $\bar Q$ is an irreducible $M$-matrix of rank $n-1$, with the nullvector $\tilde x$ associated with the simple eigenvalue at $0$, see~\cite[Section 4.3.4]{qu2009cooperative}.

\begin{thm}\label{thm:global:plane}
Consider system~\eqref{eq:x1}-\eqref{eq:x3} under Assumptions~\ref{assum:base}, \ref{assum:irreducible} and~\ref{assm:hetero:samegraph}. Further, suppose that $\rho(D^{-1}B)> 1$. 
Then
\begin{enumerate}[label=\roman*)]
    \item For all initial conditions satisfying $x^1(0) >  {\bf 0}_n$, $x^2(0) > {\bf 0}_n$, and $x^3(0) > {\bf 0}_n$, we have that $\lim_{t\to\infty} (x^1(t),x^2(t),x^3(t)) \in \mathcal{E}$ exponentially fast, where $$\mathcal {E} = \{(x^1, x^2, x^3) | \alpha_1 x^1+\alpha_2 x^2+\alpha_3 x^3 = \tilde x, \textstyle\sum_{i=1}^3 \alpha_i = 1\},$$ and $\tilde x$ is the unique endemic equilibrium of the single virus SIS dynamics defined by $(D,B)$.
    \item Every point on the connected set $\mathcal{E}$ is a coexistence equilibrium.
\end{enumerate}
\end{thm}
\textit{Proof of statement i)}: 
The initial conditions guarantee that, at some finite time $t$, we have $x^1(t) \gg  {\bf 0}_n$, $x^2(t) \gg {\bf 0}_n$, and $x^3(t)\gg {\bf 0}_n$. Define $z = x^1 + x^2 + x^3$, and $Z=X^1+X^2+X^3$.
Therefore, together with Assumption~\ref{assm:hetero:samegraph}, it follows that
\begin{align}
    \dot{z} & = \Big[-D+\big(I-(X^1(t)+X^2(t)+X^3(t))\big)B\Big] \times \nonumber \\
    &~~~~~~~~~~~~~~~~~~~~~~~~~(x^1(t)+x^2(t)+x^3(t)) \\
    & = \big[-D+(I-Z(t))B\big]z(t).\label{eq:single_virus}
\end{align}
Since $z(t) \gg {\bf 0}_n$, and $\rho(D^{-1}B) > 1$ by hypothesis, it follows from \cite[Theorem~2]{khanafer2016stability}
that $\lim_{t\to\infty} z(t) = \tilde x$ exponentially fast. In fact, $\tilde x$ is the exponentially stable equilibrium of \eqref{eq:single_virus}, with domain of attraction $z(0) \in [0,1]^n\setminus {\bf 0}$.  It follows that, for all $z(0) \in [0,1]^n\setminus {\bf 0}$, $\Vert \tilde x - z(t) \Vert \leq ae^{-bt}$ for some positive constants $a, b$, with $\Vert \cdot \Vert$ being the Euclidean norm.

The dynamics for virus~$i$, where $i\in [3]$, can be written as $$\dot{x}^i(t) = -[D+(I-\diag(\tilde x))B]x^i(t) + (\diag(\tilde x) - Z(t))Bx^i(t).$$
Without loss of generality, we consider virus~$1$ and drop the superscript. That is, we study the system
\begin{equation}\label{eq:virus_system_plane}
    \dot{x}(t) = -Qx(t) + (\diag(\tilde x) - Z(t))Bx(t),
\end{equation}
where $Z(t)$ is treated as an external time-varying input, and $Q = D-(I-\diag(\tilde x))B$ is an irreducible singular $M$-matrix, as detailed below \eqref{eq:positive_linear_x}. Define the oblique projection matrix $R = I - \tilde x \tilde u^\top$. Define also $\zeta = Rx$, and note that $\tilde u^\top \zeta = 0$ and $\zeta = {\bf 0} \Leftrightarrow x = \alpha \tilde x$ for some $\alpha \in \mathbb R$. That is, $\zeta$ is always orthogonal to $\tilde u$, and $\zeta$ is the zero vector precisely when $x$ is in the span of $\tilde x$.

Observe that $\dot \zeta(t) = R\dot{x}(t)$. Substituting in the right of \eqref{eq:virus_system_plane} for $\dot{x}(t)$, we obtain
\begin{equation}\label{eq:virus_error_system}
    \dot{\zeta}(t) = -Q\zeta(t) + R(\diag(\tilde x)-Z(t))Bx(t),
\end{equation}
by exploiting the fact that $QR = Q = RQ$. 

Consider the Lyapunov-like function
\begin{equation}\label{eq:V}
    V = \zeta(t)^\top P \zeta(t),
\end{equation}
with $P$ defined below \eqref{eq:positive_linear_x}. It is positive definite in $\zeta$. Differentiating $V$ with respect to time 
yields  \footnotesize
\begin{align}\label{eq:V_dot}
    \dot{V} & =  -2\zeta(t)^\top PQ\zeta(t) +2\zeta(t)^\top PR(\diag(\tilde x)-Z(t))Bx(t) \nonumber \\
    & = -\zeta(t)^\top \bar Q \zeta(t) +2\zeta(t)^\top PR(\diag(\tilde x)-Z(t))Bx(t).
\end{align}
\normalsize 
\seb{Due to submultiplicativity of matrix norms, $\Vert \zeta(t)\Vert \leq \Vert R \Vert \Vert x(t)\Vert$. Hence, since $\Vert x(t)\Vert$ is bounded, \eqref{eq:V_dot} yields } \footnotesize
\begin{align}
    \dot{V} & \leq -\zeta(t)^\top \bar Q \zeta(t) + \kappa \Vert \tilde x - z(t) \Vert 
    \leq -\zeta(t)^\top \bar Q \zeta(t) + \bar a e^{-b t}, \label{eq:V_dot_ineq_1}
\end{align}
\normalsize 
where $\kappa$ and $\bar a$ are positive constants, and the second inequality is due to the fact that $\Vert \tilde x - z(t) \Vert \leq ae^{-bt}$.

Let $\lambda_2$ denote the 
smallest strictly positive eigenvalue of $\bar Q$. The Courant-Fischer min-max theorem~\cite[Theorem~8.9]{zhang2011matrix} yields 
\begin{equation}\label{eq:cf_ineq}
    \frac{\zeta^\top \bar Q \zeta}{\zeta^\top \zeta} \geq \min_{\substack{v^\top \tilde u =  0\\v^\top v = 1}} v^\top \bar Q v = \lambda_2,
\end{equation}
for all $\zeta \neq {\bf 0}$ perpendicular to $\tilde u$. Since $\zeta^\top \tilde u = 0$ holds for all $\zeta \in \mathbb R$ by definition, \eqref{eq:cf_ineq} holds for all $\zeta\neq {\bf 0}$. Next, and recalling the definition of $P$ below \eqref{eq:positive_linear_x}, it follows that
\begin{equation}\label{key:ineq:1:lyap}
    \underline{p}\zeta^\top \zeta \leq \zeta^\top P \zeta \leq \bar p \zeta^\top\zeta,
\end{equation}
for all $\zeta$, where $\underline{p} = \min_{i\in [n]} \tilde u_i/\tilde x_i$ and $\bar p = \max_{i\in[n]} \tilde u_i/\tilde x_i$. 

\seb{From~\eqref{eq:V_dot_ineq_1} it follows that $\dot{V} \leq -\lambda_2 \zeta(t)^\top \zeta + \bar a e^{-bt}$, which further implies that $ \dot{V} \leq -\bar \lambda \bar p\zeta(t)^\top \zeta + \bar a e^{-bt}$, where $\bar \lambda = \lambda_2/\bar p$. Consequently, from~\eqref{key:ineq:1:lyap}, it follows that $ \dot{V} \leq -\bar\lambda V + \bar a e^{-bt}$. Thus, and recalling the definition of $\zeta$, it follows that $\lim_{t\to\infty} x(t) = \alpha \tilde x$ exponentially fast, where $\alpha \in (0,1)$ because $x(t) \in (0,1)^n$ for all $t \geq \tau$, for some positive $\tau$.}


This analysis holds not only for virus~$1$, but also virus~$2$ and virus~$3$. In other words, $\lim_{t\to\infty} x^i(t) = \alpha_i \tilde x$ for some $\alpha_i \in (0,1)$, for all $i\in [3]$. Recall that $\lim_{t\to\infty} z(t) = \tilde x$, and we immediately conclude that $\sum_{i=1}^n \alpha_i = 1$, \seb{thus proving statement i).} 

\textit{Proof of statement ii)}: We now prove that every point in $\mathcal{E}$ is an equilibrium (coexistence follows trivially by definition). Consider an arbitrary point $(x^1, x^2, x^3)$ in $\mathcal{E}$. From \eqref{eq:x1}, we have \vspace{-4mm}
\begin{align}
    \dot{x}^1(t) & = (-D+(I-X^1-X^2-X^3)B)x^1 \\
    & = (-D+(I-\diag(\tilde x))B)\alpha_1 \tilde x = {\bf 0}.
\end{align}
By the same arguments, it follows that $\dot{x}^2(t) = {\bf 0}$ and $\dot{x}^3(t) = {\bf 0}$ at the point $(x^1, x^2, x^3)$ in $\mathcal{E}$. In other words, $(x^1, x^2, x^3)$ is an equilibrium of the system system~\eqref{eq:x1}-\eqref{eq:x3}. Since this holds for any arbitrary point in $\mathcal{E}$, the \seb{proof of statement~ii) is complete.} \qed

\section{Simulations}\label{sec:simulations}

We now present a set of simulations which highlight the key theoretical results of our paper. We choose $D^i = I_n$ for $i = 1,2,3$, with the following $B^i$ matrices, where $\hat{\beta}^k_{ij}$ are constants that are changed depending on the simulation example being presented. 
\[   B^1 = \footnotesize \begin{bmatrix}
    0 & 0 & 0 & 1.5\\1.5 & 0 & 0 & 0\\0 & 1.5 & 0 & 0\\0 &0 &1.5& 0
    \end{bmatrix},\\
    B^2 = \begin{bmatrix}
    0 & 1.5+\hat{\beta}^2_{12} & 0 & 0\\0 & 0 & 1.5 & 0\\0 & 0 & 0 & 1.5\\1.5& 0& 0& 0
    \end{bmatrix},
\]
\begin{equation}
    B^3 =  \footnotesize \begin{bmatrix}
    1 & 0 & 0.5+\hat{\beta}^{3}_{13} & 0\\0 & 1+\hat{\beta}^{3}_{22} & 0.5 & 0\\0 & 0.5 & 0 & 1\\0.3+\hat{\beta}^{3}_{31}& 0 &1.2& 0
    \end{bmatrix}.\nonumber
\end{equation}
Initial conditions are obtained as follows. First, for $i\in [n]$ and $s\in[4]$, we sample a value $p_{i}^s$ from a uniform distribution $(0,1)$. Then, for $i\in [n]$ and $k\in[3]$ we set $x_{i}^k(0) = p_{ij}^k/\sum_{s=1}^4 p_{ij}^s$, which ensures that the initial conditions are in $\mathcal D$ but otherwise randomized. Note the logarithmic scale of the time axis.

\begin{figure}
\begin{minipage}{0.49\linewidth}
\centering
\subfloat[Example~1]{\includegraphics[width=\columnwidth]{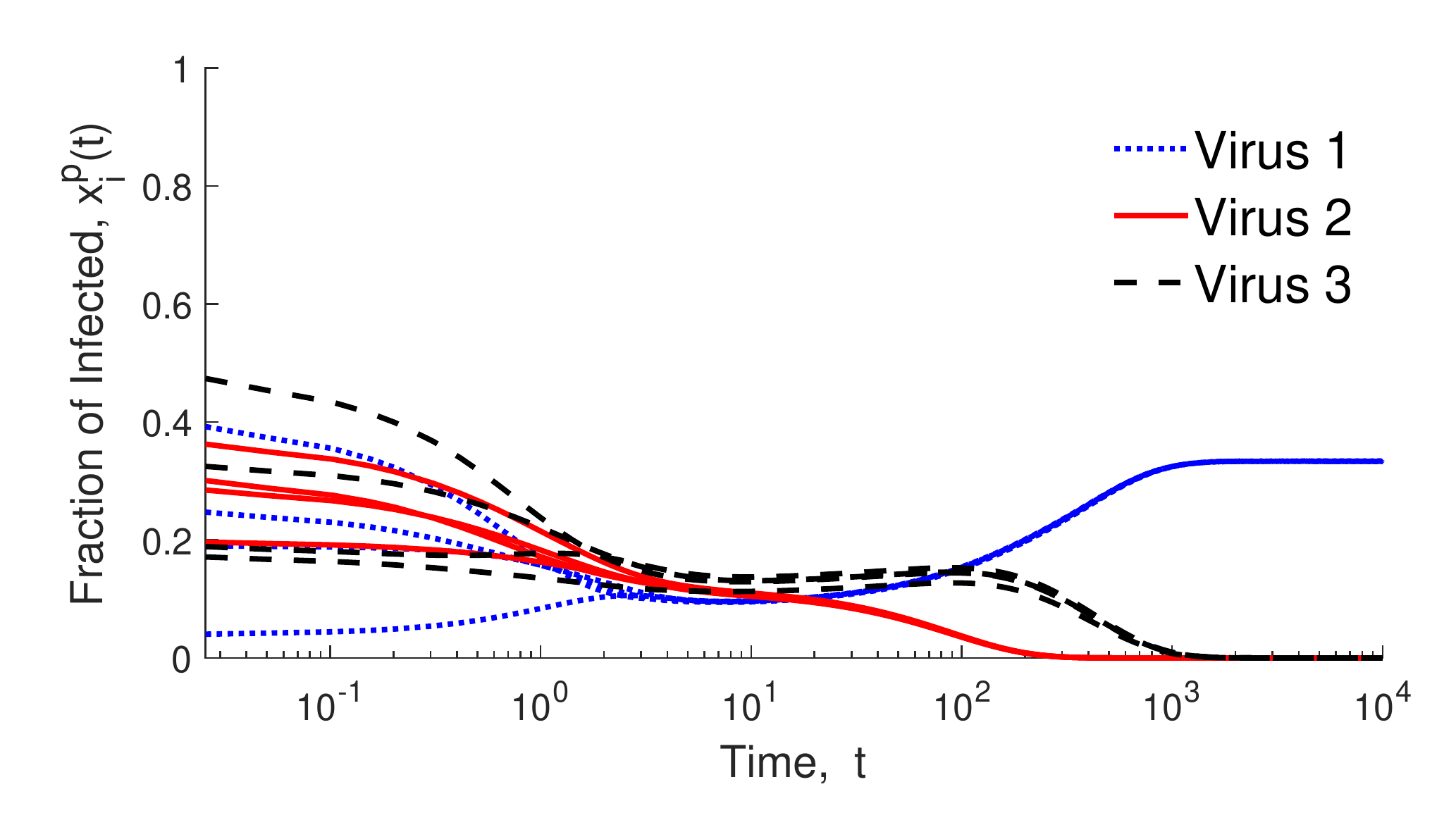}\label{fig:tri_virus_stable_virus1}}
\end{minipage}
\hfill
\begin{minipage}{0.49\linewidth}
\centering\subfloat[Example~2]{\includegraphics[width=\columnwidth]{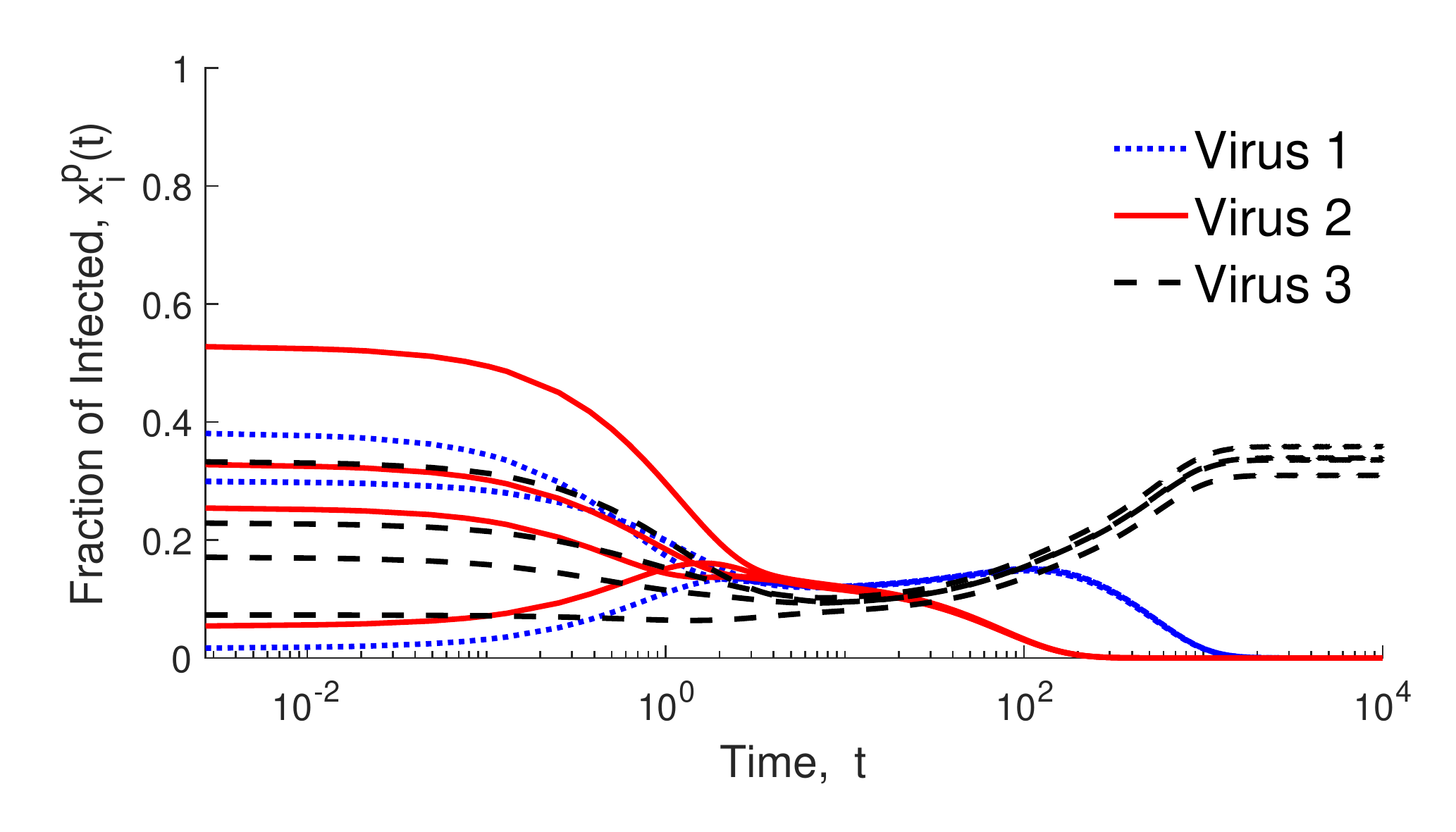}\label{fig:tri_virus_unstable_virus1}}
\end{minipage}
\vfill
\begin{minipage}{0.49\linewidth}
\centering
\subfloat[Example~3]{\includegraphics[width=\columnwidth]{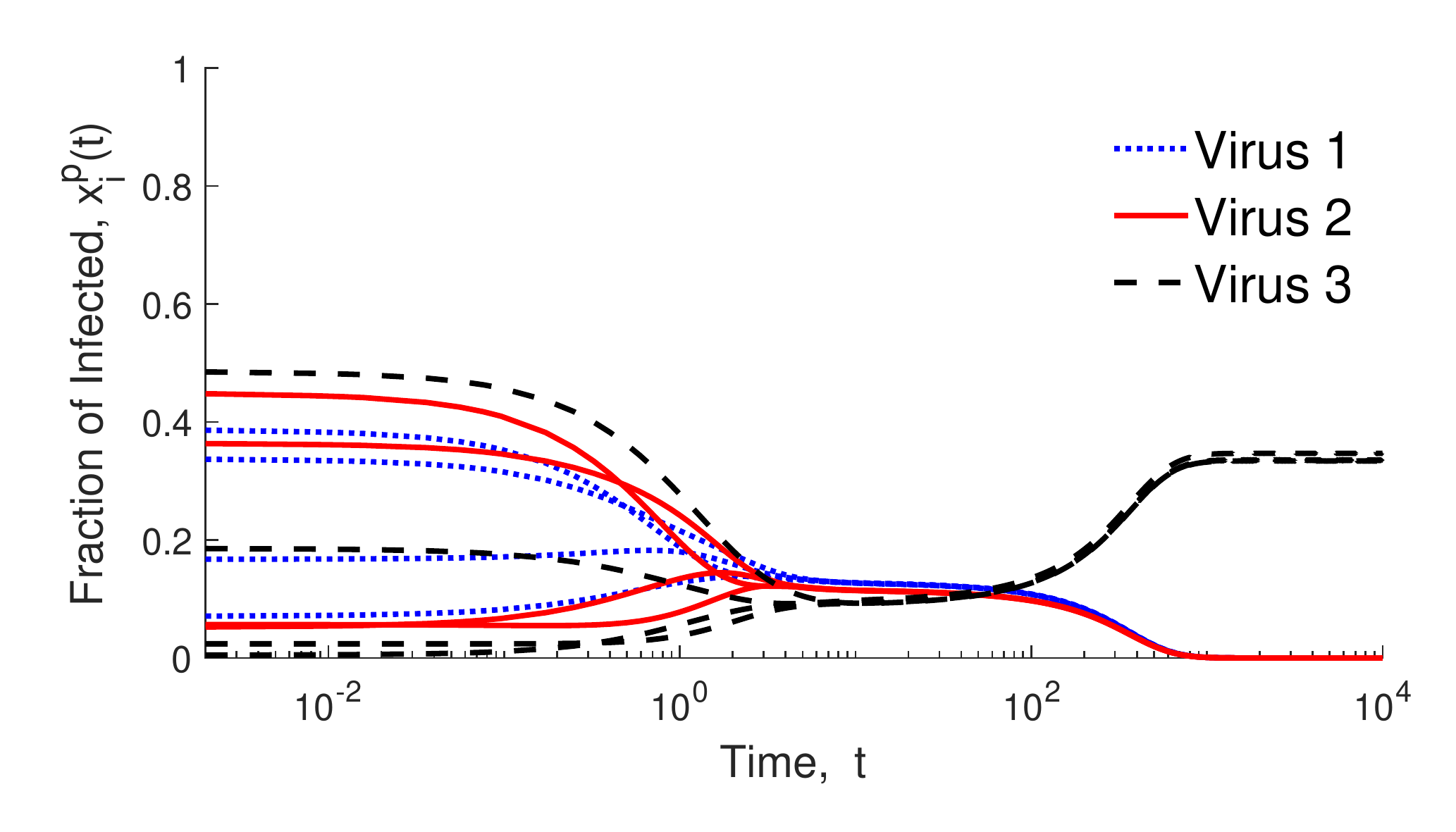}\label{fig:tri_virus_unstable_2line}}
\end{minipage}
\hfill
\begin{minipage}{0.49\linewidth}
\centering\subfloat[Example~4]{\includegraphics[width=\columnwidth]{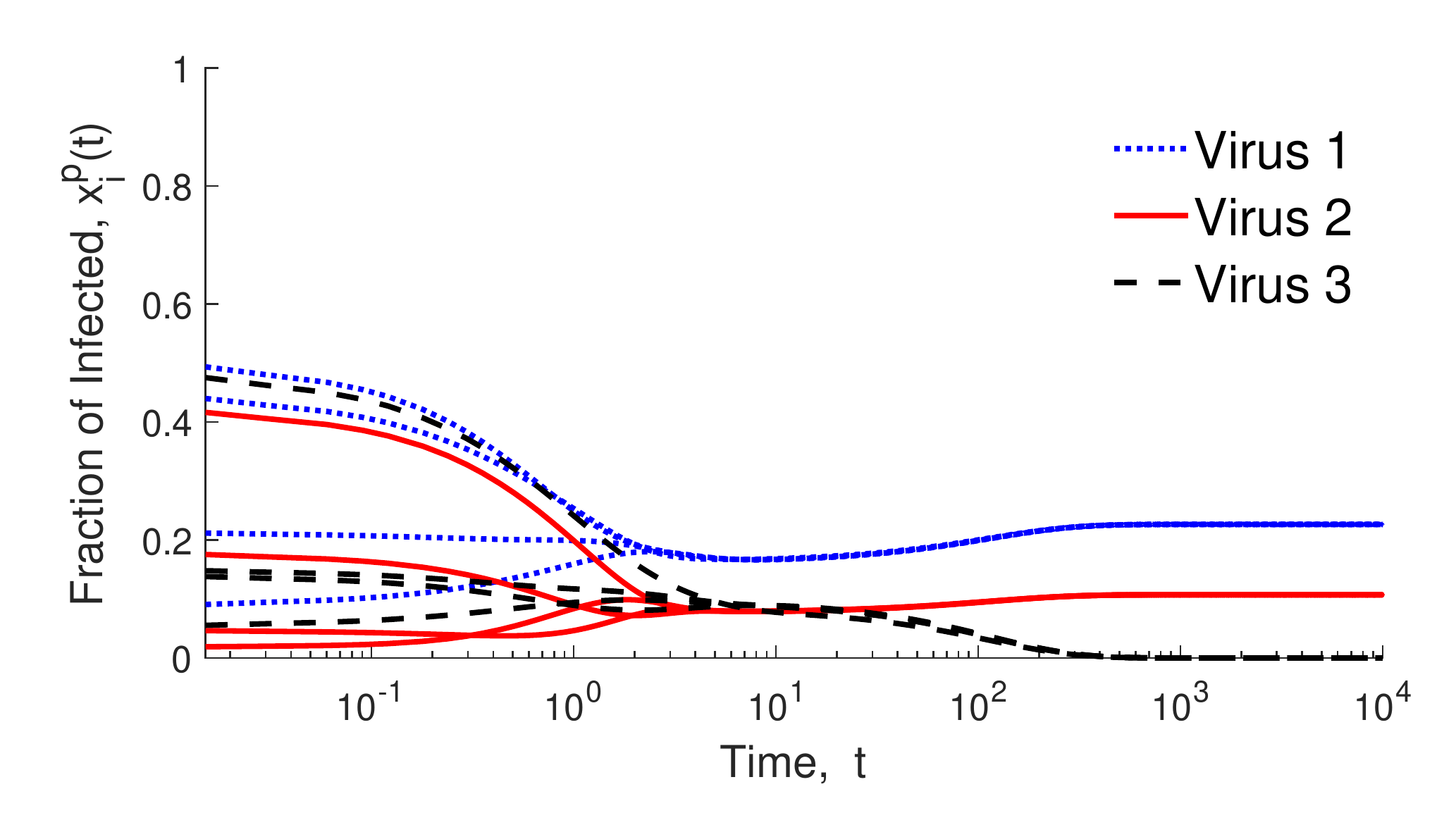}\label{fig:tri_virus_stable_2line}}
\end{minipage}
\caption{Trajectories of the simulated trivirus system \eqref{eq:full}, for different simulation parameters detailed in Section~\ref{sec:simulations}. }\label{fig:tri_virus_simulations}
\end{figure}

\textit{Example~1:} We set $\hat{\beta}^3_{13} = 0$,  $\hat{\beta}^2_{12}=-0.1$, $\hat{\beta}^3_{22} = -0.1$, and $\hat{\beta}^{3}_{31} = 0.1$. In this example, and following the notation of Proposition~\ref{prop:necessity} and Theorem~\ref{thm:local}, we obtain $\rho((I-\tilde x^1)(D^2)^{-1}B^2 = 0.9829$ and $\rho((I-\tilde x^1)(D^3)^{-1}B^3 = 0.99624$, where $\tilde X^1 = \diag(\tilde x^1)$. Thus, $(\tilde x^1, \bf{0}, \bf{0})$ is locally exponentially stable, and Fig.~\ref{fig:tri_virus_stable_virus1} shows convergence to $(\tilde x^1, \bf{0}, \bf{0})$. It can be computed that $\rho((I-\tilde x^2)(D^1)^{-1}B^1 = 1.0174$ and $\rho((I-\tilde x^2)(D^3)^{-1}B^3 = 1.0127$ and similarly, $\rho((I-\tilde x^3)(D^1)^{-1}B^1 = 1.003$ and $\rho((I-\tilde x^3)(D^2)^{-1}B^2 = 0.9863$. Hence, according to Theorem~\ref{thm:local}, $({\bf 0},\tilde x^2, {\bf 0})$ and $({\bf 0}, {\bf 0}, \tilde x^3)$ are both unstable. Additional simulations with randomized initial conditions appear to suggest that $(\tilde x^1, {\bf 0}, {\bf 0})$ is globally attractive for initial conditions in the interior of $\mathcal{D}$.

\textit{Example~2:} We set $\hat{\beta}^3_{13} = 0$,  $\hat{\beta}^2_{12}=-0.1$, $\hat{\beta}^3_{22} = -0.1$, and $\hat{\beta}^{3}_{31} = 0.15$. Consequently, we obtain $\rho((I-\tilde x^1)(D^2)^{-1}B^2 = 0.9829$ and $\rho((I-\tilde x^1)(D^3)^{-1}B^3 = 1.0037$. Indeed, we are able to compute that  $({\bf 0}, {\bf 0}, \tilde x^3)$ is locally exponentially stable according to Theorem~\ref{thm:local}, while the two other boundary equilibria are unstable; see Figure~\ref{fig:tri_virus_unstable_virus1}. Additional simulations with randomized initial conditions appear to suggest that $({\bf 0}, {\bf 0}, \tilde x^3)$ is globally attractive for initial conditions in the interior of $\mathcal{D}$.

\textit{Example~3:} We set $\hat{\beta}^3_{13} = 0.05$,  $\hat{\beta}^2_{12}=\hat{\beta}^3_{22} = \hat{\beta}^{3}_{31} = 0$. In this example, and following the notation of Theorem~\ref{thm:init:condns}, we have a line of equilibria $(\beta_1 z, (1-\beta_1)z, {\bf 0})$, with $z = \frac{1}{3}\bf{1}$ and $\beta_1\in [0,1]$. Since $\rho((I-Z)B^3 = 1.0043$, in line with statement~ii) in Theorem~\ref{thm:init:condns}, this line of equilibria is unstable; see Figure~\ref{fig:tri_virus_unstable_2line}. In fact, we can compute that $({\bf 0}, {\bf 0}, \tilde x^3)$ is locally exponentially stable, following reasoning analogous to that in Theorem~\ref{thm:local}. 
Additional simulations with randomized initial conditions appear to suggest that $({\bf 0}, {\bf 0}, \tilde x^3)$ is globally attractive for initial conditions in the interior of $\mathcal{D}$.

\textit{Example~4:} We set $\hat{\beta}^3_{13} = -0.1$,  $\hat{\beta}^2_{12}=\hat{\beta}^3_{22} = \hat{\beta}^{3}_{31} = 0$. Similar to Example~3, we have a line of equilibria $(\beta_1 z, (1-\beta_1)z, {\bf 0})$, with $z = \frac{1}{3}\bf{1}$. Now, however, $\rho((I-Z)B^3 = 0.9911$, and this line of equilibria is locally exponentially attractive according to statement i) in Theorem~\ref{thm:init:condns}; see Figure~\ref{fig:tri_virus_stable_2line}. The boundary equilibrium $({\bf 0}, {\bf 0}, \tilde x^3)$ is unstable, following reasoning analogous to that in Theorem~\ref{thm:local}. Additional simulations with randomized initial conditions appear to suggest that $(\beta_1 z, (1-\beta_1)z, {\bf 0})$ is globally attractive for initial conditions in the interior of $\mathcal{D}$, while the value of $\beta_1$ is dependent on the initial conditions.


\section{Conclusion}\label{sec:conclusion}
\seb{The present paper studied competitive tri-virus spread. In particular, it was rigorously shown that, unlike the bivirus system, the tri-virus system is not monotone. We identified a necessary and sufficient condition for local exponential convergence to a boundary equilibrium. Subsequently, a special case that admits the existence and local exponential attractivity of a continuum of coexistence equilibria was identified. Finally, we identified another special case where, irrespective of the initial non-zero infection levels, the tri-virus dynamics converge to a plane of coexistence equilibria. Thus, this paper delineated (some of) the important differences from the bivirus case 
and improved upon the findings in the existing literature.} 

\seb{As previously mentioned, no dynamical behavior can be ruled out for the tri-virus. Therefore, investigating scenarios that admit the existence (resp. exclude the possibility of existence) of limit cycles could be a line of future investigation.
Another problem of possible interest would be to understand the endemic behavior of time-varying tri-virus SIS models.}

\bibliography{ReferencesRice}

\begin{thebibliography}{10}
\providecommand{\url}[1]{#1}
\csname url@samestyle\endcsname
\providecommand{\newblock}{\relax}
\providecommand{\bibinfo}[2]{#2}
\providecommand{\BIBentrySTDinterwordspacing}{\spaceskip=0pt\relax}
\providecommand{\BIBentryALTinterwordstretchfactor}{4}
\providecommand{\BIBentryALTinterwordspacing}{\spaceskip=\fontdimen2\font plus
\BIBentryALTinterwordstretchfactor\fontdimen3\font minus
  \fontdimen4\font\relax}
\providecommand{\BIBforeignlanguage}[2]{{%
\expandafter\ifx\csname l@#1\endcsname\relax
\typeout{** WARNING: IEEEtran.bst: No hyphenation pattern has been}%
\typeout{** loaded for the language `#1'. Using the pattern for}%
\typeout{** the default language instead.}%
\else
\language=\csname l@#1\endcsname
\fi
#2}}
\providecommand{\BIBdecl}{\relax}
\BIBdecl

\bibitem{van2008virus}
P.~Van~Mieghem, J.~Omic, and R.~Kooij, ``Virus spread in networks,''
  \emph{IEEE/ACM Transactions On Networking}, vol.~17, no.~1, pp. 1--14, 2008.

\bibitem{hethcote2000mathematics}
H.~W. Hethcote, ``The mathematics of infectious diseases,'' \emph{SIAM Review},
  vol.~42, no.~4, pp. 599--653, 2000.

\bibitem{bloom2018epidemics}
D.~E. Bloom, D.~Cadarette, and J.~Sevilla, ``Epidemics and economics,''
  \emph{Finance \& Development}, vol.~55, no. 002, 2018.

\bibitem{prakash2010virus}
B.~A. Prakash, H.~Tong, N.~Valler, M.~Faloutsos, and C.~Faloutsos, ``Virus
  propagation on time-varying networks: Theory and immunization algorithms,''
  \emph{Joint European Conference on Machine Learning and Knowledge Discovery
  in Databases}, pp. 99--114, 2010.

\bibitem{lajmanovich1976deterministic}
A.~Lajmanovich and J.~A. Yorke, ``A deterministic model for gonorrhea in a
  nonhomogeneous population,'' \emph{Mathematical Biosciences}, vol.~28, no.
  3-4, pp. 221--236, 1976.

\bibitem{fall2007epidemiological}
A.~Fall, A.~Iggidr, G.~Sallet, and J.-J. Tewa, ``Epidemiological models and
  lyapunov functions,'' \emph{Mathematical Modelling of Natural Phenomena},
  vol.~2, no.~1, pp. 62--83, 2007.

\bibitem{khanafer2016stability}
A.~Khanafer, T.~Ba{\c{s}}ar, and B.~Gharesifard, ``Stability of epidemic models
  over directed graphs: A positive systems approach,'' \emph{Automatica},
  vol.~74, pp. 126--134, 2016.

\bibitem{gracy2022modeling}
S.~Gracy, P.~E. Par{\'e}, J.~Liu, H.~Sandberg, C.~L. Beck, K.~H. Johansson, and
  T.~Ba{\c{s}}ar, ``Modeling and analysis of a coupled {SIS} bi-virus model,''
  \emph{Automatica}, 2022, {N}ote: Under Review.

\bibitem{nedic2019graph}
A.~Nedi{\'c}, A.~Olshevsky, and C.~A. Uribe, ``Graph-theoretic analysis of
  belief system dynamics under logic constraints,'' \emph{Scientific reports},
  vol.~9, no.~1, pp. 1--16, 2019.

\bibitem{castillo1989epidemiological}
C.~Castillo-Chavez, H.~W. Hethcote, V.~Andreasen, S.~A. Levin, and W.~M. Liu,
  ``Epidemiological models with age structure, proportionate mixing, and
  cross-immunity,'' \emph{Journal of {M}athematical {B}iology}, vol.~27, no.~3,
  pp. 233--258, 1989.

\bibitem{carlos2}
C.~Castillo-Chavez, W.~Huang, and J.~Li, ``Competitive exclusion and
  coexistence of multiple strains in an {SIS STD} model,'' \emph{SIAM Journal
  on Applied Mathematics}, vol.~59, no.~5, pp. 1790--1811, 1999.

\bibitem{sahneh2014competitive}
F.~D. Sahneh and C.~Scoglio, ``Competitive epidemic spreading over arbitrary
  multilayer networks,'' \emph{Physical Review E}, vol.~89, no.~6, p. 062817,
  2014.

\bibitem{santos2015bi}
A.~Santos, J.~M. Moura, and J.~M. Xavier, ``Bi-virus {SIS} epidemics over
  networks: Qualitative analysis,'' \emph{IEEE Transactions on Network Science
  and Engineering}, vol.~2, no.~1, pp. 17--29, 2015.

\bibitem{liu2019analysis}
J.~Liu, P.~E. Par{\'e}, A.~Nedi{\'c}, C.~Y. Tang, C.~L. Beck, and
  T.~Ba{\c{s}}ar, ``Analysis and control of a continuous-time bi-virus model,''
  \emph{IEEE Transactions on Automatic Control}, vol.~64, no.~12, pp.
  4891--4906, 2019.

\bibitem{pare2021multi}
P.~E. Par{\'e}, J.~Liu, C.~L. Beck, A.~Nedi{\'c}, and T.~Ba{\c{s}}ar,
  ``Multi-competitive viruses over time-varying networks with mutations and
  human awareness,'' \emph{Automatica}, vol. 123, p. 109330, 2021.

\bibitem{ye2021convergence}
M.~Ye, B.~D. Anderson, and J.~Liu, ``Convergence and equilibria analysis of a
  networked bivirus epidemic model,'' \emph{SIAM Journal on Control and
  Optimization}, vol.~60, no.~2, pp. S323--S346, 2022.

\bibitem{axel2020TAC}
\BIBentryALTinterwordspacing
A.~Janson, S.~Gracy, P.~E. Par\'e, H.~Sandberg, and K.~H. Johansson,
  ``Networked multi-virus spread with a shared resource: Analysis and
  mitigation strategies,'' 2022. [Online]. Available:
  \url{https://arxiv.org/pdf/2011.07569.pdf}
\BIBentrySTDinterwordspacing

\bibitem{ben:brian:opinion:lcss}
M.~Ye and B.~D. Anderson, ``Competitive epidemic spreading over networks,''
  \emph{IEEE Control Systems Letters}, pp. 1--1, 2022.

\bibitem{pare2020analysis}
P.~E. Par{\'e}, D.~Vrabac, H.~Sandberg, and K.~H. Johansson, ``Analysis, online
  estimation, and validation of a competing virus model,'' in \emph{2020
  American Control Conference (ACC)}.\hskip 1em plus 0.5em minus 0.4em\relax
  IEEE, 2020, pp. 2556--2561.

\bibitem{pare2020modeling}
P.~E. Par{\'e}, C.~L. Beck, and T.~Ba{\c{s}}ar, ``Modeling, estimation, and
  analysis of epidemics over networks: An overview,'' \emph{Annual Reviews in
  Control}, vol.~50, pp. 345--360, 2020.

\bibitem{smith1988systems}
H.~L. Smith, ``Systems of ordinary differential equations which generate an
  order preserving flow. a survey of results,'' \emph{SIAM review}, vol.~30,
  no.~1, pp. 87--113, 1988.

\bibitem{hirsch1988stability}
M.~W. Hirsch, ``Stability and convergence in strongly monotone dynamical
  systems,'' \emph{Journal fur die reine und angewandte Mathmatik}, 1988.

\bibitem{sontag2007monotone}
E.~D. Sontag, ``Monotone and near-monotone biochemical networks,''
  \emph{Systems and synthetic biology}, vol.~1, no.~2, pp. 59--87, 2007.

\bibitem{varga1999matrix}
\BIBentryALTinterwordspacing
R.~Varga, \emph{Matrix Iterative Analysis}, ser. Springer Series in
  Computational Mathematics.\hskip 1em plus 0.5em minus 0.4em\relax Springer
  Berlin Heidelberg, 1999. [Online]. Available:
  \url{https://books.google.se/books?id=U2XYs1DyKiYC}
\BIBentrySTDinterwordspacing

\bibitem{qu2009cooperative}
Z.~Qu, \emph{Cooperative control of dynamical systems: applications to
  autonomous vehicles}.\hskip 1em plus 0.5em minus 0.4em\relax Springer Science
  \& Business Media, 2009.

\bibitem{khalil2002nonlinear}
H.~Khalil, \emph{Nonlinear Systems}.\hskip 1em plus 0.5em minus 0.4em\relax
  Prentice Hall, 2002.

\bibitem{zhang2011matrix}
F.~Zhang, \emph{{Matrix Theory: Basic Results and Techniques}}, second
  edition~ed.\hskip 1em plus 0.5em minus 0.4em\relax Springer, 2011.

\end{thebibliography}
\end{document}